\documentclass[aps,pop,twocolumn,superscriptaddress,letterpaper]{revtex4}
\usepackage{mathrsfs}
\usepackage{amsmath}
\usepackage{bm}
\usepackage{graphicx}
\usepackage{hyperref}

\begin{document}           


\title{A Full-Matrix Approach for Solving General Plasma Dispersion Relation}
\author{Hua-sheng XIE\footnote{Email: huashengxie@gmail.com}}
\affiliation{Institute for Fusion Theory and Simulation, Zhejiang
University, Hangzhou, 310027, PRC}
\date{\today}

\begin{abstract}
A hitherto difficult and unsolved issue in plasma physics is how to
give a general numerical solver for complicated plasma dispersion
relation, although we have long known the general analytical forms.
We transform the task to a full-matrix eigenvalue problem, which
allows to numerically calculate all the dispersion relation
solutions exactly free from convergence problem and give
polarizations naturally for arbitrarily complicated multi-scale
fluid plasma with arbitrary number of components. Attempt to kinetic
plasma via $N$-point Pad\'e approximation of plasma dispersion
function also shows good results.
\end{abstract}



\maketitle

Since only few simple dispersion relations are analytical tractable
in plasma physics, it is a historical issue to develop general
numerical solvers for practical applications, especially for space,
astrophysical and laser plasma. However, it is a hitherto difficulty
to develop a kinetic solver with general distribution function and
general other effects. The first problem comes from the Landau
contour integral (e.g., Landau damping) of the kinetic distribution
function. Second problem is the infinity orders of Bessel function
summation for magnetized plasma which is related to the cyclotron
resonance (e.g., Bernstein modes). These two obstacles make it
difficult to develop a root finding solver with good convergence.

WHAMP (Waves in Homogeneous Anisotropic Multicomponent Magnetized
Plasma) code by Ronnmark\cite{Ronnmark1982,Ronnmark1983} is an
important step to that goal, which can be used to calculate general
non-relativistic kinetic wave dispersion relation in plasmas with
parallel beams. To calculate fast, Pad\'e approximation is used for
plasma dispersion function. However, the numerical convergence is
still not as expectations, especially that it is difficult to give
proper initial guesses for high frequency (e.g.,
$\omega\geq10\Omega_{ci}$) modes.

Bret\cite{Bret2007} discussed how to derive and solve the 3-by-3
parallel beam-plasma dielectric tensor in fluid approximation with
relativistic effect with the help of computer ({\it Mathematica}).
However, it is still not easy to use this method if there are many
components.

In our treatment, we do not need to derive  the final 3-by-3
dispersion relation tensor matrix for $\bm E=(E_x,E_y,E_z)$ as the
conventional treatment, such as by Stix\cite{Stix1992} and used by
Ronnmark\cite{Ronnmark1982,Ronnmark1983} and Bret {\it et
al.}\cite{Bret2007,Bret2006a,Bret2006b}, which is helpful to provide
analytical insight but not a must for numerical solver.

In this manuscript (brief communication), as the first topic, we
present a trivial multi-fluid non-relativistic magnetized arbitrary
orient warm beam plasma problem, which is very difficult via usual
treatments, to show how our full-matrix approach works.

The original equations are
\begin{subequations} \label{eq:fpeq}
\begin{eqnarray}
  & \partial_t n_s = -\nabla\cdot(n_s\bm v_s),\\
  & \partial_t \bm v_s = -\bm v_s\cdot \nabla\bm v_s+\frac{q_s}{m_s}\big(\bm E+\bm v_s\times \bm B\big)-\frac{\nabla p_s}{\rho_s},\\
  & \partial_t \bm E = c^2\nabla\times\bm B - \bm J/\epsilon_0,\\
  & \partial_t \bm B = -\nabla\times\bm E,
\end{eqnarray}
\end{subequations}
with
\begin{subequations} \label{eq:fpeq2}
\begin{eqnarray}
  & \bm J = \sum_sq_sn_s\bm v_s, \\
  & d_t(p_s\rho_s^{-\gamma_s}) = 0, \label{eq:fpeq2_b}
\end{eqnarray}
\end{subequations}
where $\rho_{s}\equiv m_sn_{s}$, $p_s\equiv k_BT_{s0}/m_s$ and
$c^2=1/\mu_0\epsilon_0$.

In cold plasma limit, for parallel beam, the usual 3-by-3 dispersion
relation tensor $\overleftrightarrow{D}(\bm k,\omega)$ can be
derived as
\begin{equation}\label{eq:Dij33}
\left(\begin{array}{ccc}
    K_{xx}-n^2\cos^2\theta & K_{xx} & K_{xz}-n^2\sin\theta\cos\theta\\
    K_{yx} & K_{yy}-n^2 & K_{yz}\\
    K_{zx}-n^2\sin\theta\cos\theta & K_{zy} & K_{zz}-n^2\sin^2\theta
    \end{array}
 \right),
\end{equation}
with
\begin{eqnarray}\label{eq:drKij}
  K_{xx} &=&
  K_{yy}=1-\sum_s\frac{\omega_{ps}^2}{(\omega_s^{'2}-\omega_{cs}^2)}\Big(\frac{\omega_s'}{\omega}\Big)^2,\cr
  K_{zz} &=&
  1-\sum_s\omega_{ps}^2\bigg[\frac{1}{\omega_s^{'2}}+\frac{1}{(\omega_s^{'2}-\omega_{cs}^2)}\Big(\frac{kv_{ds}}{\omega}\Big)^2\sin^2\theta\bigg],\cr
  K_{yx} &=&
  -K_{xy}=i\sum_s\frac{\omega_{cs}\omega_{ps}^2}{\omega(\omega_s^{'2}-\omega_{cs}^2)}\frac{\omega_s'}{\omega},\cr
  K_{zx} &=&
  K_{xz}=-\sum_s\frac{\omega_{ps}^2}{(\omega_s^{'2}-\omega_{cs}^2)}\frac{\omega_s'}{\omega}\frac{kv_{ds}}{\omega}\sin\theta,\cr
  K_{zy} &=& -K_{yz}=-i\sum_s\frac{\omega_{cs}\omega_{ps}^2}{\omega(\omega_s^{'2}-\omega_{cs}^2)}\frac{kv_{ds}}{\omega}\sin\theta,
\end{eqnarray}
and $n\equiv ck/\omega$, $\omega_s'\equiv\omega-\bm k\cdot \bm
v_{s0}$, $\bm B_0=(0, 0, B_0)$, $\bm v_{s0}=(0, 0, v_{ds})$, $\bm
k=(k_x, 0, k_z)=(k\sin\theta, 0, k\cos\theta)$,
$\omega_{cs}=q_sB_0/m_s$ (note: $q_e=-e$) and
$\omega_{ps}^2=n_{s0}q_s^2/\epsilon_0m_s$. The dispersion relations
is
\begin{equation}\label{eq:dr33}
    D(\bm k,\omega)\equiv det[\overleftrightarrow{D}(\bm
    k,\omega)]=0.
\end{equation}

The summation $\sum_s$ in $K_{ij}$ ($i,j=x, y, z$) with $\omega$ in
the denominator, which is singularity at $\omega=0$ and
$\omega^{'2}=\omega_{cs}^2$, makes a general good convergence
numerical solver very difficult. And, we need also give good initial
guesses when using special root finding solver such as Newton's
iterative method or give a guess domain in complex plane using such
as Davies' method\cite{Davies1986}.

We can get an explicit polynomial form dispersion relation equation
from (\ref{eq:dr33}) for $k(\omega,\theta)$ easily. While, it is
very cumbersome to calculate an explicit form for
$\omega(k,\theta)$, even though with the aids of computer (e.g.,
using {\it Mathematica}). Without beam ($v_{ds}=0$) and for only one
ion species ($s=e, i$), a fifth order explicit form polynomial for
$\omega^2(k,\theta)$ is given in Swanson's
textbook\cite{Swanson2003}, which is simplified from hundreds of
terms.

The above investigations imply that it is not a satisfactory choice
to solve the dispersion relation using (\ref{eq:dr33}) directly as
usual treatment.

An alterant method is using the original full dispersion relation
matrix and then treating the task as a matrix eigenvalue problem,
which need neither derive the final dispersion relation equation nor
worry about how to solve it.

The linearized version of (\ref{eq:fpeq}) with
$f=f_0+f_1e^{i\bm{k\cdot r}-i\omega t}$ is equivalent to a matrix
eigenvalue problem (similar treatment can be found in
\cite{Goedbloed2004} for MHD equations)
\begin{equation}\label{eq:eig}
    \lambda\bm X=\bm M \cdot \bm X,
\end{equation}
with $\lambda=-i\omega$ the eigenvalue and corresponding $\bm X$
eigen vector, which represents the polarization information of each
normal/eigen mode solution.

For (\ref{eq:fpeq}), $\bm X$ is
\begin{equation*}\label{eq:fpdrX}
(\{n_{s1}, v_{s1x}, v_{s1y}, v_{s1z}\}, E_{1x}, E_{1y}, E_{1z},
B_{1x}, B_{1y}, B_{1z})^T,
\end{equation*}
and $\bm M$ is
\begin{widetext}
\begin{equation}\label{eq:fpdrA}
\left[\begin{array}{cccccccccc}
    \{-i\bm k\cdot\bm v_{s0} & -ik_xn_{s0} & 0 & -ik_zn_{s0} & 0 & 0 & 0 & 0 & 0 & 0\\
    -ik_xc_s^2/\rho_{s0} & -i\bm k\cdot\bm v_{s0} & \omega_{cs} & 0  & \frac{q_s}{m_s} & 0 & 0 & 0 & -\frac{q_sv_{s0z}}{m_s} & \frac{q_sv_{s0y}}{m_s}\\
    0 & -\omega_{cs} & -i\bm k\cdot\bm v_{s0} & 0  & 0 & \frac{q_s}{m_s} & 0 & \frac{q_sv_{s0z}}{m_s} & 0 & -\frac{q_sv_{s0x}}{m_s}\\
    -ik_zc_s^2/\rho_{s0} & 0 & 0 & -i\bm k\cdot\bm v_{s0}\}  & 0 & 0 & \frac{q_s}{m_s} & -\frac{q_sv_{s0y}}{m_s} & \frac{q_sv_{s0x}}{m_s} & 0\\
    -\frac{q_sv_{s0x}}{\epsilon_0} & -\frac{q_sn_{s0}}{\epsilon_0} & 0 & 0  & 0 & 0 & 0 & 0 & -ik_zc^2 & 0\\
    -\frac{q_sv_{s0y}}{\epsilon_0} & 0 & -\frac{q_sn_{s0}}{\epsilon_0} & 0  & 0 & 0 & 0 & ik_zc^2 & 0 & -ik_xc^2\\
    -\frac{q_sv_{s0z}}{\epsilon_0} & 0 & 0 & -\frac{q_sn_{s0}}{\epsilon_0}  & 0 & 0 & 0 & 0 & ik_xc^2 & 0\\
    0 & 0 & 0 & 0  & 0 & ik_z & 0 & 0 & 0 & 0\\
    0 & 0 & 0 & 0  & -ik_z & 0 & ik_x & 0 & 0 & 0\\
    0 & 0 & 0 & 0  & 0 & -ik_x & 0 & 0 & 0 & 0
    \end{array}
 \right],
\end{equation}
\end{widetext}
where we have used $p_{s1}=\gamma_sp_{s0}\rho_{s1}/\rho_{s0}$, which
is from (\ref{eq:fpeq2_b}), and $c_s^2\equiv
\gamma_sp_{s0}/\rho_{s0}$. The effects from non-zero equilibrium
quantities ($\bm J_0$ and $\bm v_{s0}\times\bm B_0$) are omitted
here (Note: This is another annoying unsolved problem in
literatures. Principally, we need treat it as inhomogeneous
problem.).

For $s$ kinds of species, the dimensions of  $\bm M$ are
$(4s+6)\times(4s+6)$. Here, we can get all the solutions of the
above system exactly (without convergence problem) via standard
matrix eigenvalue solver, e.g. function {\it eig()} in MATLAB.

\begin{table}
\caption{\label{tab:clodswanson} Comparing the cold plasma solutions
using matrix method and Swanson's polynomial. Only positive
solutions are shown, since $\omega_{+}^2=\omega_{-}^2$.}
\begin{ruledtabular}
\begin{tabular}{cccc}
  $\omega^M$ & $\omega^S$ & $\omega^M$ & $\omega^S$ \\\hline
  10.5152 & 10.5152 & 1.1330E-4 & 1.1330E-4+i1E-16 \\
  10.0031 & 10.0031 & - & 1E-32+i1E-18 \\
  9.5158 & 9.5158 & - & 0 \\
  2.4020E-4 & 2.4020E-4-i3E-17 &  &  \\
\end{tabular}
\end{ruledtabular}
\end{table}

Cold limit ($p_{s0}=0$), without beam ($\bm v_{s0}=0$), $s=e,i$,
numerical solutions of (\ref{eq:eig}) ($\omega^M$) and Swanson's
polynomial ($\omega^S$) are given in Table \ref{tab:clodswanson},
with $kc=0.1$, $\theta=\pi/3$, $m_i/m_e=1836$ and
$\omega_{pe}=10\omega_{ce}$. The results are consistent with each
other exactly, except some small ($<10^{-15}$) numerical errors.

\begin{figure}
  \includegraphics[width=8cm]{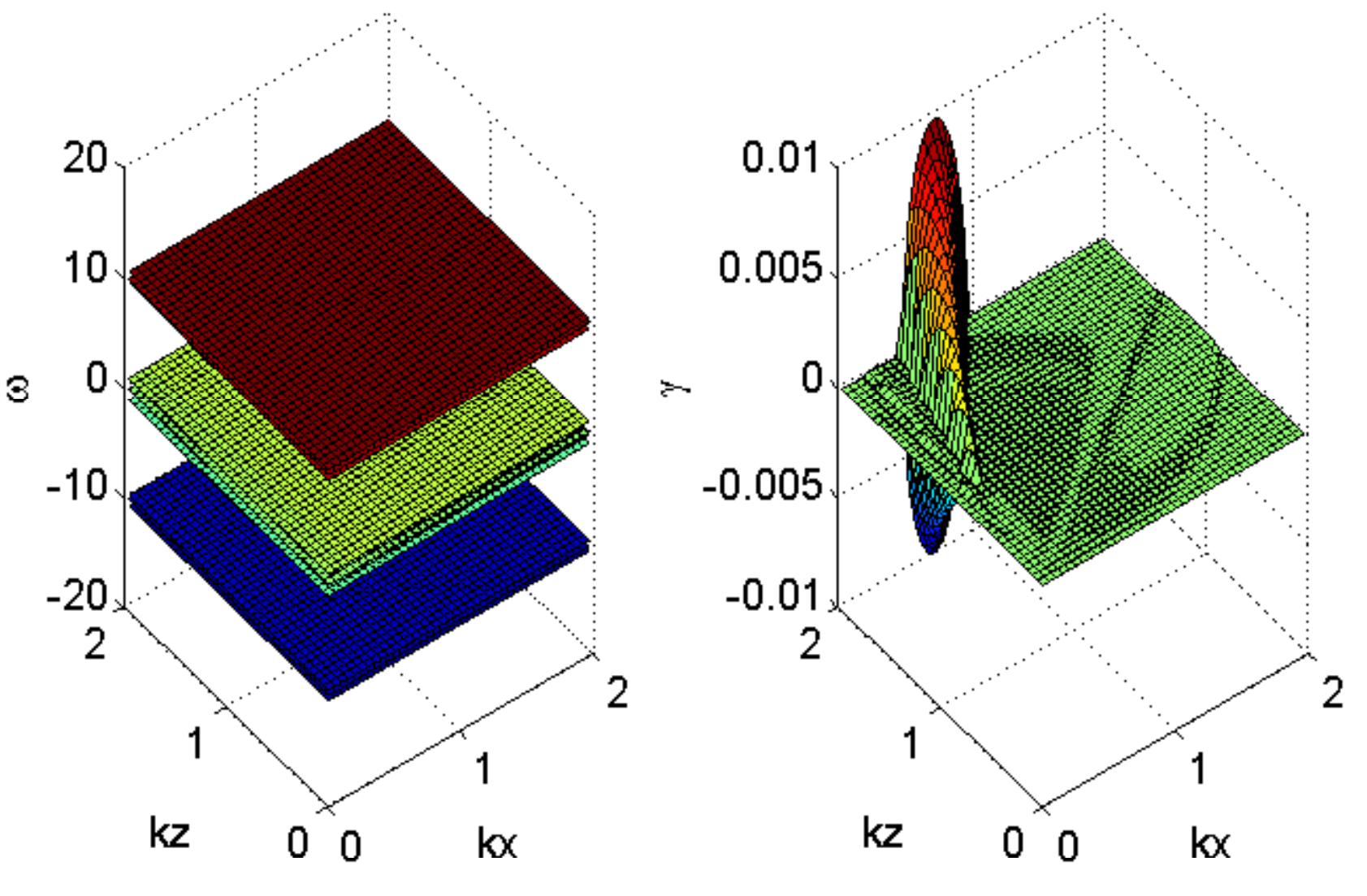}\\
  \caption{Test run for fluid beam plasma, with $c_{1e,2e,i}=0.01c$, $v_{1ex}=0.01c$, $n_{1e}=0.9$, $v_{2ex}=-0.09c$ and $n_{2e}=0.1$.}\label{fig:fluid_beam}
\end{figure}

A 3-component result with perpendicular electron beams is shown in
Fig.\ref{fig:fluid_beam}.

Using this method for other arbitrarily complicated fluid system is
just straightforward, due to that it is just a directly rewriting of
original linearized equations without any approximations.

As a second topic, we try this matrix method for kinetic plasma.

A possible general method is discreting\cite{Bratanov2012, Xie2013}
the distribution function in $v$-space or using basis function
expansion\cite{Ng1999} to transform the equations to matrix form.
However, it is shown that for both Vlasov-Possion\cite{Bratanov2012}
and Vlasov-Ampere\cite{Xie2013} systems, the (Landau) damping
($\Im(\omega)<0$) normal modes are not eigenmodes but just related
to spectral density accumulating of eigenmodes, though those methods
can give some correct solutions for growth modes. Another problem is
that the matrix dimension should be very large to calculate
accurately since discrete of $v$ is introduced.

Another attempt is the initial value version\cite{Xie2012,Xie2013}
of this matrix method, which can give correct kinetic Landau
damping, but gives only few largest imaginary part modes and can not
overcome multi-scale problem.

Here, as our first successful attempt to kinetic plasma, we limit to
simple multi-component 1D electrostatic (ES1D) problem with drift
Maxwellian distribution, with dispersion relation
\begin{equation}\label{eq:drkes1d}
    D=1-\sum_s\frac{1}{(k\lambda_{Ds})^2}\frac{Z'(\zeta_s)}{2}=0,
\end{equation}
where $\lambda_{Ds}^2=\frac{\epsilon_0k_BT_s}{n_sq_s^2}$,
$v_{ts}=\sqrt{\frac{2k_BT_s}{m_s}}$ and
$\zeta_s=\frac{\omega-kv_{s0}}{kv_{ts}}$, and use $N$-point Pad\'e
approximation of plasma dispersion function as by Ronnmark
\cite{Ronnmark1982,Ronnmark1983}
\begin{equation}\label{eq:NZ}
    Z'(s)=\sum_j\frac{b_j}{(s-c_j)^2},
\end{equation}
where $N=8$ is used, which shows to be very accurate for most domain
in upper plane (except $y<\sqrt{\pi}x^2e^{-x^2}$, $x\gg1$, with
$s=x+iy$). Bad performances are just for strong damping domain, for
which we have little interest.

Combining (\ref{eq:drkes1d}) and (\ref{eq:NZ}), gives a very similar
dispersion relation equation
\begin{equation}\label{eq:drkes1d2}
    1-\sum_s\sum_j\frac{b_{sj}}{(\omega-c_{sj})^2}=0,
\end{equation}
as the one of the fluid ES1D beam plasma, with
$b_{sj}=\frac{b_jv_{ts}^2}{2\lambda_{Ds}^2}$ and
$c_{sj}=k(v_{s0}+v_{ts}c_j)$, which can help us transform the
problem to an equivalent linear system
\begin{subequations} \label{eq:fpeq}
\begin{eqnarray}
  & \omega A_{sj} = b_{sj}B_{sj}+c_{sj}A_{sj},\\
  & \omega B_{sj} = c_{sj}B_{sj}+C,\\
  & C = \sum_{sj}A_{sj},
\end{eqnarray}
\end{subequations}
which is an eigenvalue prolem of a $2sN\times 2sN$ dimensions eigen
matrix $\bm M$, with $sN=s\times N$. The sigularity of dominator,
which meet in conventional treatment, is canceled after this
transformation. And the matrix method can support multi-component
very easily and naturally.

For Langmuir wave Landau damping, calculating the largest imaginary
part solution using matrix method ($\omega^M$) and original
$Z(\zeta)$ function ($\omega^Z$) are shown in Table \ref{tab:LD}.

\begin{table}
\caption{\label{tab:LD} Comparing the Landau damping solutions using
matrix method and original $Z(\zeta)$ function.}
\begin{ruledtabular}
\begin{tabular}{ccccc}
  $k\lambda_{De}$ & $\omega_r^M$ & $\omega_i^M$ & $\omega_r^Z$ & $\omega_i^Z$ \\\hline
  0.1 & 1.0152 & 0.0000 & 1.0152 & -4.8E-15 \\
  0.5 & 1.4157 & -0.1533 & 1.4157 & -0.1534 \\
  1.0 & 2.0458 & -0.8513 & 2.0458 & -0.8513 \\
  2.0 & 3.1897 & -2.8278 & 3.1891 & -2.8272 \\
\end{tabular}
\end{ruledtabular}
\end{table}

\begin{figure}
  \includegraphics[width=8cm]{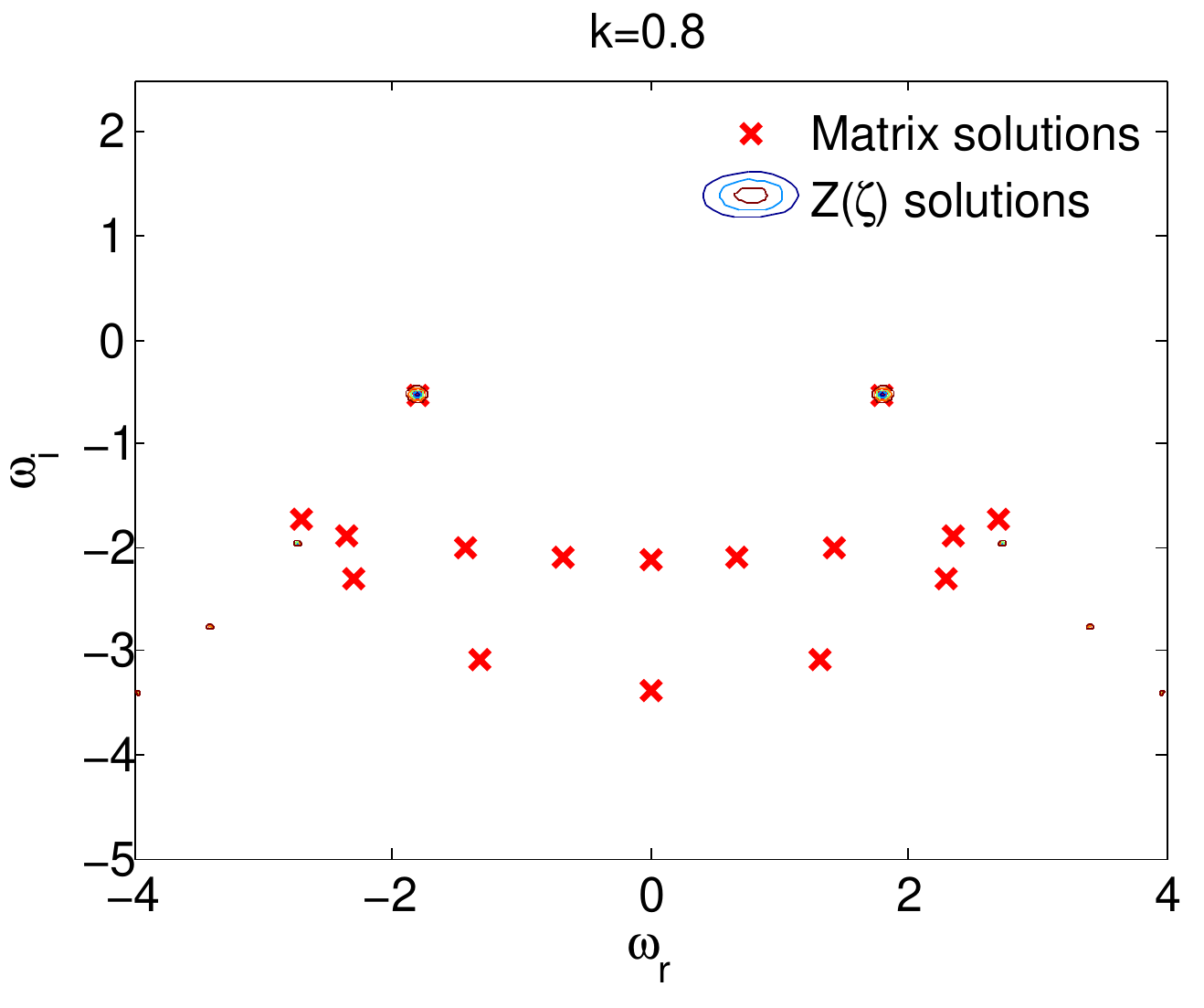}\\
  \caption{Comparing of all solutions from matrix method and $Z(\zeta)$ function}\label{fig:pdrk_landau}
\end{figure}

Fig. \ref{fig:pdrk_landau} shows all the solutions of matrix method
and also solutions using $Z(\zeta)$ function for $k=0.8$. We can see
that only the largest imaginary part solutions are overlapping,
which is enough for what we want.

For two-frequency-scale ion acoustic mode, besides the Langmuir mode
$\omega=2.0458-0.8513i$, the largest imaginary part solution in
matrix method is also consistent with $Z(\zeta)$ function solution,
e.g., $T_i=T_e$, $m_i=1836m_e$, $k\lambda_{De}=1$, gives
$\omega=0.0420-0.0269i$.

\begin{figure}
  \includegraphics[width=8cm]{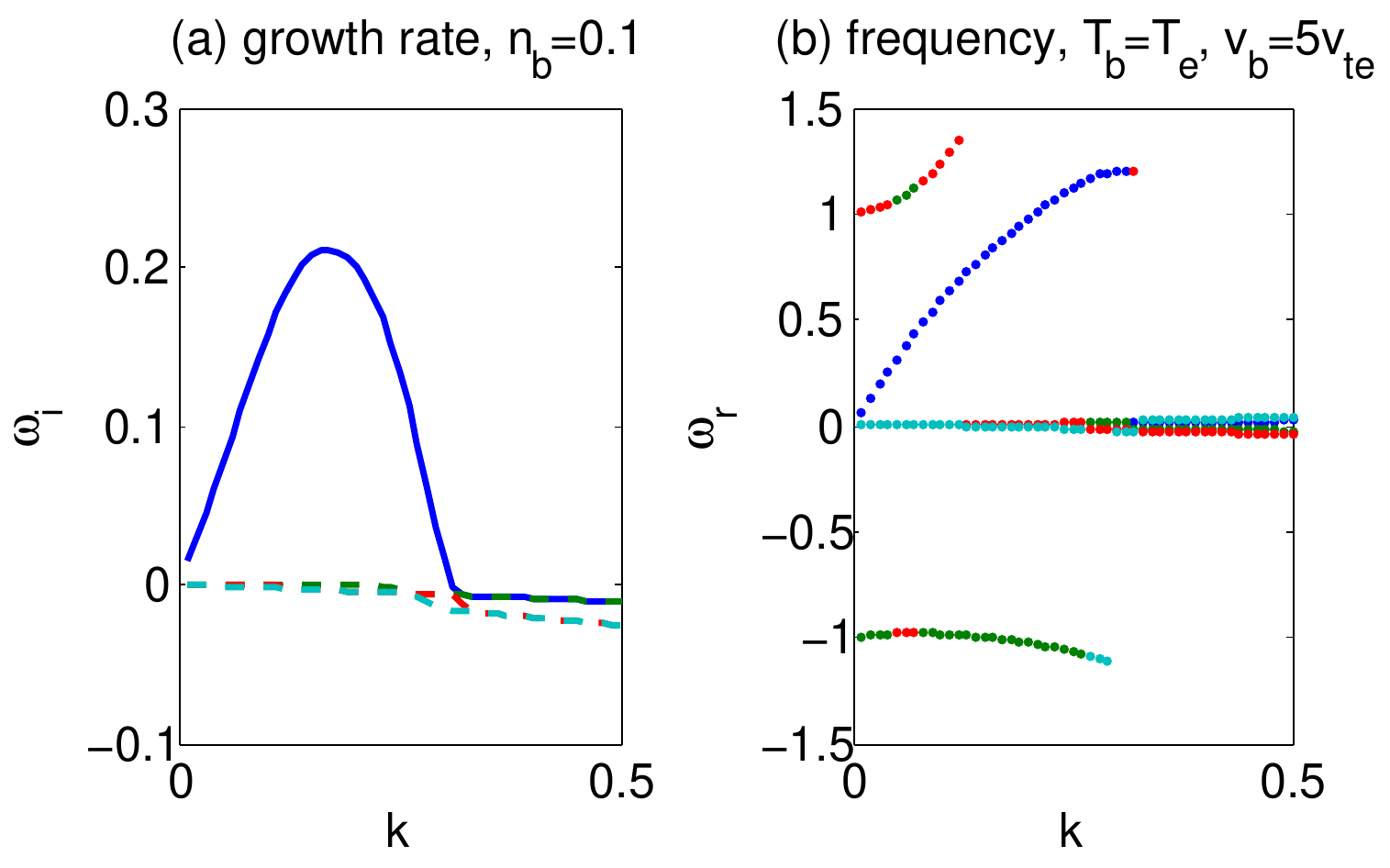}\\
  \caption{ES1D electron bump-on-tail modes with ion effect.}\label{fig:pdrk_beam}
\end{figure}

First four largest imaginary part solutions of electron bump-on-tail
mode, with ion effects, are shown in Fig.\ref{fig:pdrk_beam} for
$T_b=T_e=T_i$, $v_b=5v_{te}$ and $n_b=0.1$. For this test run,
matrix method calculates very fast and automatically, and can give
all the solutions we want. While, using $Z(\zeta)$ function, we need
test different initial guess one by one to select different modes
(not shown here).

Detail features and benchmarks of highly nontrivial solvers for
multi-component magnetized and un-magnetized fluid plasma
$\lambda\bm B\bm X=\bm A\bm X$ with and without anisotropic,
relativistic, beam and gradient effects, will be reported elsewhere,
where $\bm B\neq\bm I$ is mainly from relativistic factor. Though
haven't shown yet, we hope, in the future, this method can also
performances well for other nontrivial kinetic problems.

In summary, a very efficacious method is presented to solve both
fluid and (simple) kinetic plasma dispersion relation, which have
overcome many (almost all) troubles in conventional treatments.

[The author would like to thank M. Y. Yu for improving the
manuscript.] Conversations with Liu Chen, Y. Xiao and other
researchers at IFTS-ZJU are appreciated. Information from Y. Lin, X.
Y. Wang, R. Denton and Ling Chen are also acknowledged.

Codes for solving (\ref{eq:eig}) and (\ref{eq:fpeq}) are provided as
supplementary materials for one who hopes to quickly access to this
matrix approach.


\begin{thebibliography}{99}

\bibitem{Ronnmark1982} K. Ronnmark, WHAMP - Waves
in Homogeneous Anisotropic Multicomponent Magnetized Plasma, KGI
Report No. 179, Sweden, 1982.

\bibitem{Ronnmark1983} K. Ronnmark, Computation of the
dielectric tensor of a Maxwellian plasma, Plasma Physics, 1983, 25,
699.

\bibitem{Bret2007} A. Bret, Beam-plasma dielectric tensor with Mathematica, Computer
Physics Communications, 2007, 176, 362 - 366.

\bibitem{Bret2006a} A. Bret and C. Deutsch, A fluid approach to linear beam plasma
electromagnetic instabilities, Physics of Plasmas, 2006, 13, 042106.

\bibitem{Bret2006b} A. Bret, M. E. Dieckmann and C. Deutsch, Oblique electromagnetic
instabilities for a hot relativistic beam interacting with a hot and
magnetized plasma, Physics of Plasmas, 2006, 13, 082109.

\bibitem{Stix1992} T. Stix, \emph{Waves in Plasmas} (AIP, New York, 1992).

\bibitem{Davies1986} B. Davies, Locating the zeros of an analytic function, Journal of
Computational Physics, 1986, 66, 36 - 49.


\bibitem{Goedbloed2004} J. Goedbloed and S. Poedts, \emph{Principles of Magnetohydrodynamics:
With Applications to Laboratory and Astrophysical Plasmas}
(Cambridge, 2004).

\bibitem{Swanson2003} D. G. Swanson, \emph{Plasma
Waves} (IOP, 2nd Ed., 2003).

\bibitem{Bratanov2012} V. Bratanov, ``Landau and Van Kampen Spectra in Discrete Kinetic Plasma Systems", Master thesis, LMU
Munich,
\url{http://www.theorie.physik.uni-muenchen.de/TMP/theses/thesis\_bratanov.pdf},
2012.

\bibitem{Ng1999} C. S. Ng, A. Bhattacharjee and F. Skiff, Phys. Rev. Lett.
\textbf{83}, 1974 (1999).

\bibitem{Xie2013} H. S. Xie, A $1/t$ damped electrostatic electron plasma
wave, submitted to PoP.

\bibitem{Xie2012} H. S. Xie, arXiv:1211.5984.


\end{thebibliography}
\end{document}